\documentclass[aps,prd,twocolumn,superscriptaddress,nofootinbib,10pt]{revtex4-1}   

\pdfoutput=1
\usepackage[utf8]{inputenc}
\usepackage{amssymb}
\usepackage{amsmath}
\usepackage{amsfonts}
\usepackage{graphicx}
\usepackage{color}
\usepackage{xspace}
\usepackage{comment}
\usepackage{hyperref}
\usepackage[normalem]{ulem}

\usepackage[section]{placeins}
\usepackage{afterpage}

\usepackage{float}
\usepackage{slashed}
\usepackage{ulem}
\usepackage{appendix}

\usepackage{multirow,rotating}
\usepackage[dvipsnames]{xcolor}

\usepackage{orcidlink}

\begin{document}
	
        \title{Long-lived sterile neutrinos from an axionlike particle at Belle II}

   \author{Zeren Simon Wang\,\orcidlink{0000-0002-1483-6314}}
   \email{wzs@mx.nthu.edu.tw}
   \affiliation{School of Physics, Hefei University of Technology, Hefei 230601, China}

   \author{Yu Zhang\,\orcidlink{0000-0001-9415-8252}}
   \email{dayu@hfut.edu.cn \,(Corresponding author)}
   \affiliation{School of Physics, Hefei University of Technology, Hefei 230601, China}

   \author{Wei Liu\,\orcidlink{0000-0002-3803-0446}}
   \email{wei.liu@njust.edu.cn \,(Corresponding author)}
   \affiliation{Department of Applied Physics and MIIT Key Laboratory of Semiconductor Microstructure and Quantum Sensing, Nanjing University of Science and Technology, Nanjing 210094, China}

\begin{abstract}
Axionlike particles (ALPs) can be produced in meson decays via tree-level quark-flavor-violating couplings, and decay to a pair of sterile neutrinos if also coupled to them. Such light sterile neutrinos are necessarily long-lived and may give rise to striking signatures of displaced vertices (DVs) at terrestrial experiments. In this work, we study the prospect of the ongoing Belle II experiment for detecting the sterile neutrinos from the ALP, leveraging the $\mathcal{O}(10^{10})$ $B$-mesons projected to be produced at the experiment with an integrated luminosity of 50 ab$^{-1}$. We propose search strategies for one or two DVs, and perform Monte-Carlo simulations in order to estimate the sensitivity reach of Belle II to active-sterile-neutrino mixing $|V_{eN}|^2$ as functions of the sterile-neutrino mass. Signatures include a selected list of sterile-neutrino final states, for which we estimate an $\mathcal{O}(1)$ or lower background level. Our results show that the proposed search strategies can probe values of $|V_{eN}|^2$ up to about two orders of magnitude beyond the existing bounds, for ALP mass over 4 GeV and sterile-neutrino mass above the $D$-meson threshold. Compared to the one-DV search, the two-DV one, despite its weaker sensitivities as a result of double exponential suppression of the sterile-neutrino decay positions, possesses the advantage of possible full reconstruction of the signal event allowing for pinning down the masses of the sterile neutrino and the ALP, if a discovery is made.
\end{abstract}


	\vskip10mm
	
	\maketitle
	\flushbottom

\section{Introduction}\label{sec:intro}

In 2012, the Large Hadron Collider (LHC) at CERN announced the discovery of a Standard-Model (SM)-like Higgs boson, of mass around 125 GeV~\cite{ATLAS:2012yve,CMS:2012qbp}.
Huge amount of effort has since been paid in order to find any new fundamental particle predicted in various beyond-the-Standard-Model (BSM) scenarios.
However, the attempts in this direction have been unsuccessful so far; while hints of BSM physics in the form of anomalies have been reported, no concrete discovery has been made yet.
At the same time, the lower bounds on the masses of e.g.~scalar quarks predicted in supersymmetry-inspired models (see Refs.~\cite{Nilles:1983ge,Martin:1997ns} for reviews) have been raised to a couple of TeV~\cite{ParticleDataGroup:2024cfk}.
This situation has in recent years led the community to wonder if the underlying assumptions made on the form of BSM physics so far are not aligned with the Nature.
One such assumption is the new physics manifesting itself in terms of heavy new particles that decay promptly once produced.
This concept has been prevalent such that the LHC has mainly searched for BSM-physics signatures as prompt objects with large transverse momentum or missing energy.
Given that new physics has eluded the LHC searches to date, another search direction has become increasingly more important, i.e.~signatures arising from long-lived particles (LLPs) that have a relatively long lifetime such that they decay only after travelling some macroscopic distance.
Although such signatures have been searched for in the past, it is only during the recent years that it has gained more attention~\cite{Alimena:2019zri,Lee:2018pag,Curtin:2018mvb,Beacham:2019nyx}.

Numerous BSM theories predict LLPs naturally.
These range from ``portal-physics'' models (see Ref.~\cite{Antel:2023hkf} for an overview of the current status), R-parity-violating supersymmetry~\cite{Barbier:2004ez,Dreiner:1997uz,Mohapatra:2015fua}, to ``neutral-naturalness'' models~\cite{Chacko:2005pe,Burdman:2006tz,Craig:2015pha,Cai:2008au,Cohen:2018mgv,Cheng:2018gvu} and so on.
These models are strongly motivated for various reasons including explaining non-vanishing neutrino masses and the existence of DM, and solving the hierarchy problem.
In this work, we will focus on a model predicting both an axionlike particle (ALP) and sterile neutrinos.
The axion is a hypothetical particle that appears after the spontaneous breaking of a global $U(1)_{PQ}$ symmetry~\cite{Peccei:1977hh} proposed for solving the strong CP problem.
It is a pseudoscalar particle with its mass related to the breaking scale of the $U(1)_{PQ}$ symmetry.
The ALP is not required to obey this relation, but rather has them as independent parameters.
Thus, it does not necessarily solve the strong CP problem as the axion does, but offers richer phenomenologies at experiments~\cite{Bauer:2017ris}.

The ALP is predicted in various ultra-violet (UV) complete models including string compactifications~\cite{Cicoli:2013ana,Ringwald:2012cu}, Froggatt-Nielsen models of flavor~\cite{Froggatt:1978nt,Alanne:2018fns}, and supersymmetry models~\cite{Bellazzini:2017neg}.
The ALP can couple to all the particles in the SM spectrum, and in this work, we study the ALP coupled only to the SM quarks off-diagonally.
Such quark-flavor-violating (QFV) couplings can arise in UV-complete models including DFSZ- and KSVZ-type axion models~\cite{Ema:2016ops,Calibbi:2016hwq,Arias-Aragon:2017eww,Bjorkeroth:2018ipq,delaVega:2021ugs,DiLuzio:2023ndz,DiLuzio:2017ogq,Alonso-Alvarez:2023wig}, among others~\cite{Gavela:2019wzg,Bauer:2020jbp,Bauer:2021mvw,Chakraborty:2021wda,Bertholet:2021hjl}.
In particular, in some of these models the quark-flavor-diagonal couplings are suppressed, such as the astrophobic axion~\cite{DiLuzio:2017ogq} where the axion in DFSZ-like models is assigned family-dependent PQ charges so that the axion couplings to the first-generation quarks are suppressed and QFV couplings arise.
Several Froggatt-Nielsen model variants also impose suppression of certain diagonal couplings and generation of QFV couplings~\cite{delaVega:2021ugs,Calibbi:2016hwq,Linster:2018avp}.
In this phenomenological work, we will restrict ourselves to the single coupling of the ALP with the bottom and strange quarks.
We choose to remain agnostic about the theoretical origin of this flavor structure of the ALP with the quarks, and treat this coupling as an independent parameter.
Past studies on QFV couplings of the ALP can be found in e.g.~Refs.~\cite{Gorbunov:2000ht,MartinCamalich:2020dfe,Carmona:2021seb,Bauer:2021mvw,Carmona:2022jid,Beltran:2023nli,Li:2024thq,Cheung:2024qve}.

Moreover, the ALP can couple to a pair of sterile neutrinos~\cite{Berryman:2017twh,Carvajal:2017gjj,Alves:2019xpc,Gola:2021abm,deGiorgi:2022oks,Abdullahi:2023gdj,Cataldi:2024bcs,Marcos:2024yfm,Wang:2024mrc} which are SM-singlet fermion fields and originally proposed for explaining the non-vanishing active-neutrino masses via the type-I seesaw mechanism~\cite{Minkowski:1977sc,Yanagida:1979as,Mohapatra:1979ia,Gell-Mann:1979vob,Schechter:1980gr}.
In the type-I seesaw model, the sterile neutrinos are predicted to be as massive as $10^{15}$ GeV close to the Grand-Unified-Theory scale, for $\mathcal{O}(1)$ Yukawa couplings.
Such a heavy sterile neutrino is beyond the reach of the present and near-future experiments.
However, there exist low-scale seesaw mechanisms~\cite{Mohapatra:1986aw,Mohapatra:1986bd,Akhmedov:1995ip,Malinsky:2005bi} where such constraint can be loosened, and much lighter, e.g.~GeV-scale, sterile neutrinos are allowed.
Sterile neutrinos in this mass range are often called heavy neutral leptons~\cite{Shrock:1980vy,Shrock:1980ct,Shrock:1981wq} and are possible to probe or even find at the current and upcoming experiments.
In this work, for kinematic reasons, the sterile-neutrino mass is restricted to be below half the assumed mass of the ALP as we will focus on an on-shell produced ALP.
Furthermore, in numerical analysis of this work, we will fix the ALP mass at some representative values: 4.2 and 4.7 GeV, since, as we will numerically find out, that only for ALP mass above 4 GeV we can probe sterile neutrinos of mass above the $D$-meson threshold and exclude model parameter space beyond the present bounds; we note that the choice of 4.7 GeV is almost just below the kinematic threshold $m_{B^\pm}-m_{K^\pm}\approx 4.79$ GeV.
Moreover, we assume the sterile neutrino is of Majorana nature.

In phenomenological studies on sterile neutrinos of this mass regime, the decoupling of their masses and mixing angles with the active neutrinos is often assumed, and we will follow the practice.
In addition, we will work in a simplified model, where there is only one kinematically relevant sterile neutrino $N$, which mixes with the electron neutrino dominantly, thus leaving only the mass $m_N$ and the mixing parameter $|V_{eN}|^2$ in the sterile-neutrino sector, besides the ALP-sterile-neutrino coupling.
The sterile neutrino can then decay via the SM weak currents allowed by non-vanishing $|V_{eN}|^2$, and for the GeV-scale sterile neutrino, the current bounds on the mixing parameter render $N$ necessarily long-lived.

In fact, not only have the main experiments at the LHC performed multiple searches for LLPs (see e.g.~Refs.~\cite{CMS:2023mny,CMS:2024trg,ATLAS:2023oti}), but also future detectors as dedicated experiments for LLP searches have been proposed to be installed in the vicinity of various LHC interactions points (IP) such as FASER~\cite{Feng:2017uoz,FASER:2018eoc}, CODEX-b~\cite{Gligorov:2017nwh,Aielli:2019ivi}, and MATHUSLA~\cite{Chou:2016lxi,Curtin:2018mvb,MATHUSLA:2020uve}.
In particular, FASER has been approved, constructed, and launched, with initial results published in Refs.~\cite{FASER:2023zcr,FASER:2023tle,FASER:2024hoe}.
Besides the LHC, other types of terrestrial experiments such as beam-dump or fixed-target experiments, and $B$-factories, can explore LLPs as new physics as well.
For instance, in March 2024, the beam-dump experiment, Search for Hidden Particles (SHiP)~\cite{SHiP:2015vad,Alekhin:2015byh,SHiP:2018xqw,SHiP:2021nfo,Albanese:2878604} has been officially approved.
In this work, we will focus on the $B$-factory experiment Belle II~\cite{Belle-II:2010dht,Belle-II:2018jsg}, which is currently under operation.
At Belle II, electron and positron beams collide asymmetrically at the center-of-mass (COM) energy of the $\Upsilon(4S)$ resonance (10.58 GeV) leading to as many as $5.5\times 10^{10}$ $B$-meson pair production events expected with its planned total integrated luminosity of 50 ab$^{-1}$.
Such huge numbers of $B$-mesons produced allow for the ALP production in $B\to K$ transitions via the QFV coupling of the ALP we mentioned above, on which we will focus in this study. See e.g.~Refs.~\cite{Berezhiani:1989fs,Berezhiani:1990wn,Berezhiani:1990jj,Ferber:2022rsf} for previous studies on ALP physics in $B$-meson decays.

The lifetime of the ALP depends on the masses of the ALP and the sterile neutrino, as well as its coupling to a pair of the sterile neutrinos.
Thus, the ALP decays into a pair of long-lived sterile neutrinos, either promptly or after traversing a macroscopic distance.
In this work, we will focus on the case of a short-lived ALP.
The sterile neutrinos may subsequently undergo displaced decays inside the Belle II detector leading to spectacular signatures of displaced vertices (DVs) which suffer from little to no background at the tracking system of Belle II.
Phenomenological studies on LLP searches at Belle II can be found in Refs.~\cite{Zhou:2021ylt,Dey:2020juy,Filimonova:2019tuy,Dib:2019tuj,Ferber:2022rsf,Bertholet:2021hjl,Cheung:2021mol,Kim:2019xqj,Kang:2021oes,Acevedo:2021wiq,Dreyer:2021aqd,Duerr:2019dmv,Duerr:2020muu,Chen:2020bok,Bandyopadhyay:2022klg,Guadagnoli:2021fcj,Dib:2022ppx,Cheung:2024oxh,Ferber:2022ewf,Schafer:2022shi,Jaeckel:2023huy}.
See also Refs.~\cite{Belle-II:2023ueh,Belle:2024wyk} for recent results of LLP searches reported from Belle or Belle II.
We note that for similar signal processes in the same model, recently Ref.~\cite{Wang:2024mrc} has studied the discovery potential of the LHC far detectors and the SHiP experiment.

This work is structured as follows.
In Sec.~\ref{sec:model_constraints} we detail the theoretical model we consider and discuss the existing bounds.
We then introduce the Belle II experiment and propose our search strategies in Sec.~\ref{sec:experiment_strategies}.
Sec.~\ref{sec:simulation_computation} is dedicated to explaining the procedures of numerical simulation and computation of the signal-event rates.
We present and discuss our numerical results in Sec.~\ref{sec:results}.
Finally, we provide a summary and an outlook in Sec.~\ref{sec:conclusions}.

\section{Theoretical model and existing constraints}\label{sec:model_constraints}

We consider a low-energy effective Lagrangian of the ALP, containing terms describing its interactions with the SM quarks~\cite{Bauer:2017ris,Bauer:2018uxu,Bauer:2021mvw,Carmona:2021seb,Beltran:2023nli} and the sterile neutrinos~\cite{Alves:2019xpc,Gola:2021abm,Marcos:2024yfm}, in addition to the kinetic and mass terms,
\begin{eqnarray}
\mathcal{L}_{a} &=& \frac{1}{2} \, \partial_\mu a  \,\partial^\mu a - \frac{1}{2}m_a^2 a^2  + \frac{\partial_\mu a}{\Lambda}  \sum_q \sum_{i,j} g^q_{i,j} \bar{q_i}\gamma^\mu q_j\nonumber\\
&&+ \frac{\partial_\mu a}{\,\Lambda}\, g_N   \overline{N}  \gamma^\mu \gamma_5  N,
\label{eqn:Lagrangian_ALP}
\end{eqnarray}
where $a$ labels the ALP with mass $m_a$, $\Lambda$ is the effective cutoff scale, $q_{i,j}$ denotes the quarks with $q$ summing over $u_L, u_R, d_L, d_R$ and $i,j=1, 2, 3$ being the generation indices.

The $g^q_{i,j}$ and $g_N$ couplings are dimensionless.
For $g^q_{i,j}$ with unequal $i$ and $j$, meson decays in the form $P\to P'/V a$ may be induced where $P$ and $P'$ denote pseudoscalar mesons and $V$ a vector meson.
Here, we focus on the flavor indices $(3,2)$ in the down-type quark sector.
In particular, for the couplings $g^d_{3,2}=g_{3,2}^{d_R}+g_{3,2}^{d_L}$ and $g^d_{3,2}=g_{3,2}^{d_R}-g_{3,2}^{d_L}$, we have various $B$-meson decays into a pseudoscalar or a vector meson, plus an ALP.
Concretely, with $g^d_{3,2}=g_{3,2}^{d_R}+g_{3,2}^{d_L}$ we have the $B^+\to K^+ a$ decay along with $B^0\to K^0 a$ and $B_s^0\to \eta/\eta' a$.
The coupling $g^d_{3,2}=g_{3,2}^{d_R}-g_{3,2}^{d_L}$, on the other hand, leads to the $B^0\to K^{*0}, B^+\to K^{*+}$, and $B^0_s\to \phi$ transitions.
Higher resonances of the kaons may also contribute to ALP production in $B$-meson decays~\cite{DallaValleGarcia:2023xhh}.
For simplicity, we assume that either $g_{3,2}^{d_L}$ or $g_{3,2}^{d_R}$ is zero, so that a single coupling controls both $B\to P'$ and $B\to V$ transitions.
In the rest of the paper, we will simply use $g^d_{3,2}$ to denote the coupling inducing the $B^+\to K^+ a$ decay, and ignore the other $b\to s$ transitions at the hadron level listed above.
We note that the charge-conjugated channel $B^-\to K^- a$ is also included in the numerical study of this work.
All other ALP couplings with the quarks are assumed to be zero.

To compute the decay branching ratio of $B^+\to K^+ a$, we use the following formula~\cite{Beltran:2023nli}: 
\begin{eqnarray}
 \Gamma\left(B^+ \to K^+ a\right) &=& \frac{|g^d_{3,2}|^2}{64\pi\Lambda^2}
 \left|F_0^{B^+ \to K^+}(m_a^2)\right|^2 m_{B^+}^3 \nonumber \\
 &&\left(1 - \frac{m_{K^+}^2}{m_{B^+}^2}\right)^2
 \lambda^{1/2}\left(\frac{m_{K^+}^2}{m_{B^+}^2}, \frac{m_a^2}{m_{B^+}^2}\right)\,,\,\,\,\,\,\,\,\,\label{eqn:GammaBp2Kpa} 
\end{eqnarray}
where $F_0^{B^+\to K^+}$ is the transition form factor defined in Ref.~\cite{Wirbel:1985ji} and extracted from Ref.~\cite{FlavourLatticeAveragingGroupFLAG:2021npn}, $\lambda(x,y)\equiv 1+x^2+y^2-2\,x -2\,y -2xy$, and $m_{B^+} (m_{K^+})$ denotes the mass of the $B^+$ ($K^+$)-meson.

In this work, since we study very long-lived sterile neutrinos, the ALP essentially decays invisibly with a branching ratio of 100\%.
This allows us to derive bounds on Br$(B^+ \to K^+ a)$ and hence $g^d_{3,2}/\Lambda$ from measurements on Br$(B^+\to K^+ \nu \bar{\nu})$; note that $g^d_{3,2}/\Lambda$ should also be constrained from upper bounds on Br$(B^{+/0}\to K^{*+/0}\nu\bar{\nu})$ via Br$(B^{+/0}\to K^{*+/0} a)$.
We first consider the latest measurement on Br$(B^+\to K^+ \nu \bar{\nu})$ performed at Belle II~\cite{Belle-II:2023esi}, which reported a value Br$(B^+\to  K^+ \nu\bar{\nu})_{\text{exp}}=(2.3\pm 0.7)\times 10^{-5}$ that is larger than the SM prediction~\cite{Altmannshofer:2009ma,Buras:2014fpa}, Br$(B^+\to K^+\nu  \bar{\nu})_{\text{SM}}=(4.29\pm 0.23)\times 10^{-6}$, by 2.7$\sigma$.
To invoke non-interfering new physics (NP) beyond the SM to explain this anomaly (without taking into account the distribution of the invariant mass squared $(q^2)$ of the missing energy), the NP contribution is thus required to lie within the window Br$(B^+\to K^+\nu\bar{\nu})_{\text{NP}}=(1.9\pm 0.7)\times 10^{-5}$~\cite{He:2023bnk}.
Rather than \textit{explain} the excess, we focus on the parameter regions of $(m_a, g^d_{3,2}/\Lambda)$ that are \textit{allowed} by the experimental measurement.
Concretely, for the ALP masses we will assume for numerical analysis, i.e.~4.2 and 4.7 GeV, instead of performing a detailed bin analysis on $q^2$, we fix $g^d_{3,2}/\Lambda$ at $1\times 10^{-9}$ GeV$^{-1}$.
They correspond to Br$(B^+\to K^+ a)\sim 2.3\times 10^{-7}$ and $1.3\times 10^{-7}$, respectively, both well below the $2\sigma$ lower bound of the NP window $(1.9\pm 0.7)\times 10^{-5}$ quoted above.
Moreover, the leading experimental upper bounds on Br$(B^+\to K^{*+} \nu \bar{\nu})\lesssim 4.0\times 10^{-5}$~\cite{Belle:2013tnz} and Br$(B^0\to K^{*0}\nu \bar{\nu})\lesssim 1.8\times 10^{-5}$~\cite{Belle:2017oht} lead to upper limits on NP contributions to these branching ratios at $3.1\times 10^{-5}$ and $1.0\times 10^{-5}$, respectively~\cite{He:2022ljo}.
We have computed the decay rates of $B^{+/0}\to K^{*+/0}a$ with formulas given in Ref.~\cite{Beltran:2023nli} and found that our benchmark values of $m_a$ with the fixed coupling $g^d_{3,2}/\Lambda=1\times 10^{-9}$ GeV$^{-1}$ are all allowed by these bounds.

Before proceeding, we comment on the impact if non-zero quark-flavor-diagonal couplings of the ALP that have strengths similar to that of the considered QFV coupling are present.
In this case, they can induce contributions to the QFV coupling at loop level that would modify the production rate of the ALP, and can also affect the decay rates of the ALP.
The loop-level contributions to the ALP production in $B^+ \to K^+ a$ are negligible, compared to those from the tree-level QFV coupling.
As for modifying the ALP decay rates, since in the present paper we will restrict ourselves to a relatively large coupling of the ALP with the sterile neutrinos (that satisfies the present bound; see below), we have numerically checked and verified that the partial decay widths of the ALP induced by the quark-flavor-diagonal couplings would be orders of magnitude smaller than that of the ALP into a pair of sterile neutrinos and therefore also negligible.

The decay of the ALP proceeds via the single coupling $g_N$, into a pair of Majorana sterile neutrinos: $a\to N N$, and the corresponding decay width is computed with~\cite{Alves:2019xpc}:
\begin{eqnarray}
    \Gamma(a\to N N)=\frac{1}{2\pi} \Big(\frac{g_N}{\Lambda}\Big)^2  m_N^2 \, m_a \sqrt{1-\frac{4 m_N^2}{m_a^2}}.\label{eqn:width_alp}
\end{eqnarray}
In this work, we fix $g_N/\Lambda$ at $10^{-3}$ GeV$^{-1}$ as a benchmark value, considering perturbativity requirement $g_N m_N/\Lambda<1$.
The decay width of the ALP is saturated by $\Gamma(a\to N N)$, and for this value of $g_N/\Lambda$ and the ranges of $m_a$ and $m_N$ we are interested in, the ALP is promptly decaying unless $m_N$ is either almost zero, or extremely close to the kinematic threshold $m_a/2$.

We also note that if $g_N/\Lambda$ is orders of magnitude smaller than $10^{-3}$ GeV$^{-1}$, the ALP would become long-lived and the expected sensitivities should be weaker than those derived in the present work; an illustration of this effect can be found in Ref.~\cite{Wang:2024mrc}.
In another word, our results also hold for other values of $g_N/\Lambda$ as long as the perturbativity is obeyed and the ALP is promptly decaying at Belle II.
The kinematically relevant upper reach of $m_N$ in this study is $m_N< (m_{B^+}-m_{K^+})/2 \approx 2.4$ GeV.

For the sterile neutrino, besides the coupling with the ALP, it also participates in the SM electroweak interactions via its tiny mixing with the electron neutrino.
The Lagrangian for the sterile-neutrino interactions with the SM $W$- and $Z$-bosons after electroweak symmetry breaking is given below,
\begin{eqnarray}
		\mathcal{L}_{N} &=& \frac{g}{\sqrt{2}}\ \sum_{\alpha}
		V_{\alpha N}\ \bar \ell_\alpha \gamma^{\mu} P_L N W^-_{L \mu}\nonumber\\
		&&+		\frac{g}{2 \cos\theta_W}\ \sum_{\alpha, i}V^{L}_{\alpha i} V_{\alpha N}^* \overline{N} \gamma^{\mu} P_L \nu_{i} Z_{\mu},
		\label{eqn:Lagrangian_N}
\end{eqnarray}
where $g$ is the SM SU(2) gauge coupling, $V_{\alpha N}$ is the mixing parameters between the sterile neutrino and the active neutrinos of flavor $\alpha=e, \mu, \tau$, $\theta_W$ is the Weinberg angle, and $V^L_{\alpha i}$ is the mixing matrix between the flavor and mass eigenstates of the active neutrinos with $i=1, 2, 3$.
We note that in the present paper we consider the single $N$ mixed with the electron neutrino only and therefore for $V_{\alpha N}$ only $V_{eN}$ is non-zero.

For the computation of the decay widths of the sterile neutrino at tree level, we make use of the formulas provided in Refs.~\cite{Atre:2009rg,DeVries:2020jbs,Bondarenko:2018ptm}.
In the next section, we will discuss in detail the collider signal final states from sterile-neutrino decays we consider and the corresponding potential background sources.

The sterile neutrinos have been constrained from various experiments.
The current leading bounds on $|V_{eN}|^2$ for $m_N$ between 0.1 GeV and 2.4 GeV were obtained at beam-dump and fixed-target experiments~\cite{PIENU:2017wbj,Bryman:2019bjg,CHARM:1985nku,NA62:2020mcv,T2K:2019jwa,Barouki:2022bkt}, as well as at the LHC~\cite{CMS:2024ake}.
They are in the order of $\mathcal{O}(10^{-10}\text{--}10^{-7})$ depending on $m_N$.
These bounds were obtained for the sterile neutrino in the minimal scenario, where both the production and decay of the sterile neutrino are mediated by the mixing parameter $|V_{eN}^2|$ only.
Since our theoretical scenario concerns only $B$-meson decays for the production of the sterile neutrinos, the bounds obtained in processes where the sterile neutrinos are considered to be produced in other particles' decays still hold.
This is the case with all the leading bounds on $|V_{eN}|^2$ for the relevant sterile-neutrino mass range.
These constraints will be displayed in the numerical results of this work.
For $2.0 \text{ GeV}\lesssim m_N\lesssim 2.4$ GeV, a sub-leading bound stems from a Belle search~\cite{Belle:2013ytx} where long-lived sterile neutrinos are produced from $B$-meson decays; however, the search requires a prompt lepton which is missing in our signal process, and hence cannot place corresponding bounds.
Therefore, we conclude that the extracted present bounds are all valid in our theoretical scenario.

\begin{figure}[t]
	\centering
	\includegraphics[width=0.495\textwidth]{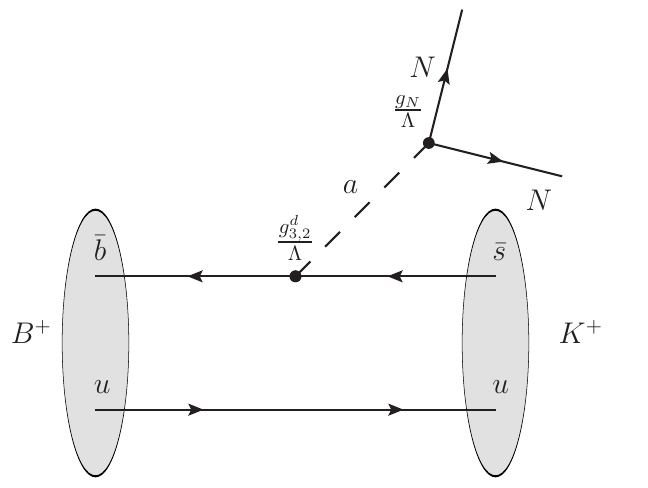}
	\caption{Feynman diagram for the pair production of the sterile neutrinos from ALP decays via $g_N/\Lambda$, with the on-shell ALP $a$ produced in the $B^+\to K^+$ transition mediated by the coupling $g^d_{3,2}/\Lambda$.
 The charge-conjugated channel is also included in the numerical study.
 }
 \label{fig:feynman_hnl}
\end{figure}

Finally, for clarity of presentation, we show in Fig.~\ref{fig:feynman_hnl} a Feynman diagram depicting the signal process of the pair production of the sterile neutrinos in the decays $B^+\to K^+ a, a\to N N$.
The mediating couplings $g^d_{3,2}/\Lambda$ and $g_N/\Lambda$ are labeled at the corresponding vertices.
We emphasize that we restrict ourselves to the case of an on-shell and promptly decaying ALP.

\section{The Belle II experiment and search strategies}\label{sec:experiment_strategies}

The Belle II experiment is an asymmetrical detector located at the interaction point (IP) of the SuperKEKB collider, Tsukuba, Japan.
With an electron beam of energy 7 GeV and a positron beam of energy 4 GeV colliding, the collider achieves a COM energy of 10.58 GeV lying at the $\Upsilon(4S)$ resonance.
The resonance decays to a $B^+ B^-$ pair with 51.4\% probability~\cite{ParticleDataGroup:2024cfk}, resulting in copious production rates of these $B$-mesons.
In total, $2.827\times 10^{10}$ $B^+ B^-$ pair-production events are expected, for the planned integrated luminosity of 50 ab$^{-1}$ by 2027 at Belle II.

The Belle II detector is composed of various subdetector components.
In this work, since we will propose search strategies based on tracking capabilities, we take the vertex detector and the tracker (Central Drift Chamber (CDC)) as the fiducial volume.
In detail, we consider $-40<z<120$ cm and $1\text{ or }10< r< 80$ cm for the definition of the fiducial volume, where $z$ and $r$ are the longitudinal and transverse coordinates of the DV, respectively.
The cutoff of $r> 1$ or 10 cm is imposed in order to facilitate suppression of background events from neutral-kaon decays, prompt tracks originating at the IP, as well as particle interactions with the materials~\cite{Dib:2019tuj,Dey:2020juy,Bertholet:2021hjl,Cheung:2021mol,Cheung:2024oxh,Belle:2024wyk}.
The other requirements on $r$ and $z$ are determined according to the geometry of the CDC, leaving at least 40 cm from its edges to ensure accurate tracking~\cite{Belle-II:2010dht}.
The choice of $r>1$ cm or $r>10$ cm depends on the applied search strategy, as explained below.

The sterile neutrino may decay into various possible SM final states via its mixing with the electron neutrino as considered here, for which the reconstruction efficiency of a DV and the expected background level at a detector differ distinctly.
We focus on sterile-neutrino decays into at least two tracks, as potential signature candidates, in this work, because such final states have good efficiencies of DV reconstruction at the Belle~II experiment.
These potential signal final states can be classified into two categories:
\begin{enumerate}
    \item leptonic $l l' \nu$: $e^\mp \mu^\pm \nu_{\mu}$, $e^\mp \tau^\pm \nu_{\tau}$, $e^- e^+\nu_e$, $\mu^- \mu^+\nu_e$,
    \item semi-leptonic  $e^\pm M^\mp$: $e^\pm \pi^\mp$, $e^\pm \rho^\mp$, $e^\pm K^\mp$, $e^\pm K^{*\mp}$, $e^\pm D^\mp$, $e^\pm D_s^\mp$,  $e^\pm D^{*\mp}$, $e^\pm D_s^{*\mp}$,
\end{enumerate}
where $l$ and $l'$ label charged leptons, and $M^\pm$ denotes a charged meson.
The specific choice of the particular signal final states depends on the search strategy to be introduced below.

In our signal process, at most two DVs can be observed in each signal event.
Therefore, we propose two search strategies requiring the reconstruction of exactly one or two DVs, called ``1DV'' and ``2DV'', respectively.
Typically the existence of a DV already suppresses background events.
Therefore, with the 2DV search, the extra DV compared to the 1DV search further lowers the background level.
Consequently, for the 1DV (2DV) search strategy, we require $r>10$ cm ($r>1$ cm~\cite{Cheung:2024oxh}).

Here, the primary remaining background is photon conversion~\cite{Jaeckel:2023huy}.
For the 2DV search, given the double DV suppression, we expect that only the $e^- e^+\nu_e$ channel may be mildly contaminated from such background events.
This background suppression has been discussed and studied in, e.g.~Refs.~\cite{Belle:2013ytx,BaBar:2015jvu,Lee:2018pag,Dib:2019tuj,Dey:2020juy,Cheung:2024oxh}.
Therefore, among all the potential final states with two tracks listed above, we exclude only this channel and treat all the other final states as signature of the sterile-neutrino decays.

For the 1DV signal region, the photon-conversion background can easily be serious even if we take $r>10$ cm~\cite{Jaeckel:2023huy}.
Therefore, among the leptonic channels, we discard the ones involving either a pair of electrons, a pair of muons, or an electron-muon pair.
We are thus left with the $e^\mp \tau^\pm \nu_{\tau}$ channel for which we require that the $\tau$-lepton should decay into at least three charged pions plus anything, to remove the photon-conversion background; for this we take Br$(\tau\to 3 \text{ prongs})=14.55\%$~\cite{ParticleDataGroup:2024cfk}.

For the semi-leptonic final states, we should take into account the possibility of an electron faking a pion.
Therefore, we do not include the decay channel $N \to e^\pm \pi^\mp$.
We keep the $N\to e^\pm \rho^\mp$ and $N\to e^\pm K^\mp$ channels where the $\rho^\pm$ meson is reconstructed by its decay to $\pi^\pm (\pi^0\to \gamma\gamma)$ and the charged kaon is long-lived leaving a track in the detector.
The charged kaon can be faked by an electron and thus suffer from the photon-conversion background.
However, in future real data analysis, some selection criteria can be taken in order to ensure a sufficiently low fake rate and suppression of background.
For instance, the polar angle of the tracks can be required to point into the calorimeter so that the tracks can be ensured not to induce an electromagnetic shower.
Also, tight cuts on $dE/dx$ and on signatures from the time-of-propagation detector can be applied.
We take a benchmark value of $0.1\%$ for the $e$-to-$K$ fake rate, under the above-mentioned selection criteria. 
Moreover, for these two channels, the sterile neutrinos can be fully reconstructed.
Given the mass resolution of a few MeV at Belle II~\cite{Belle-II:2018jsg} and the narrow width of the sterile neutrino, we can select a narrow window around the considered sterile-neutrino mass, which, combined with the low fake rate, largely suppress the background level from photon conversion, down to the order of $\mathcal{O}(1)$ or even less, considering the order of magnitude $\mathcal{O}(10^3)$ such background events for 50 ab$^{-1}$ integrated luminosity as estimated in Ref.~\cite{Jaeckel:2023huy}.

Finally, for the semi-leptonic channels with a $D$-meson, we, for simplicity, discard those with an excited $D$-meson which contribute very little anyway and retain only $N \to e^\pm D^\mp$ and $N\to e^\pm D_s^\mp$.
For these final states, we require that the $D$-mesons should decay into at least three charged pions to remove the photon-conversion background.

In real analysis, further requirements may be imposed to better contain the background without impairing the signal-event yield to a tangible extent, such as requiring the angle between the tracks from the DV be small.
We do not go into more detail and refer the reader to, for instance, Refs.~\cite{Dib:2019tuj,Dey:2020juy,Cheung:2024oxh,Belle:2024wyk}.
Based on the discussion above, we will assume vanishing background and show in Sec.~\ref{sec:results} contour curves of 3 signal events as the sensitivity reach at $95\%$ confidence level (CL.).

\begin{figure}[t]
	\centering
	\includegraphics[width=0.495\textwidth]{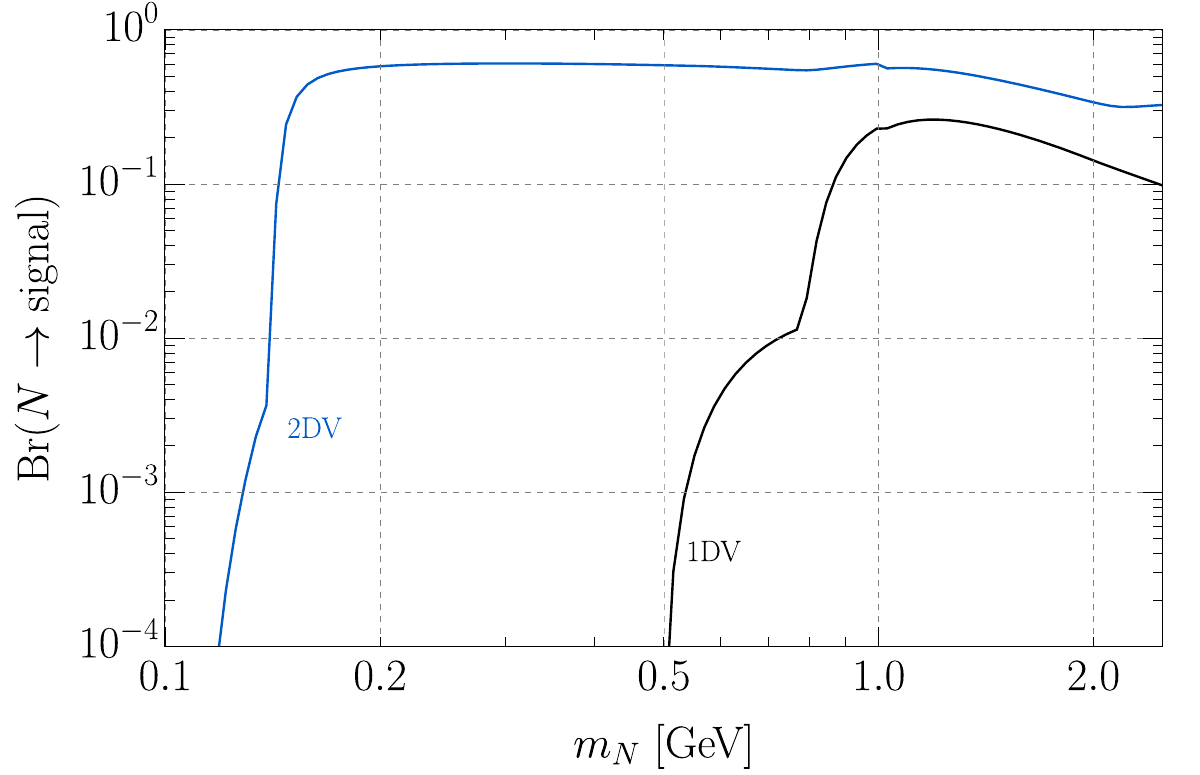}
	\caption{Decay branching ratio of $N$ into signal final states, as functions of $m_N$, for the 1DV (black) and 2DV (blue) search strategies, respectively.
 }
 \label{fig:BrNtoVisible}
\end{figure}
In Fig.~\ref{fig:BrNtoVisible}, we present a plot of the decay branching ratios of the sterile neutrino $N$ into the signal final states,Br$(N\to\text{signal})$, as functions of the sterile-neutrino mass $m_N$, for both 1DV and 2DV search strategies separately.
In the 2DV curve we observe a kink at the pion mass where the $e^\pm \pi^\mp$ channel opens up.
The 1DV curve starts at the kaon threshold because the lowest signal decay channel is $N\to e^\pm K^\mp$, and its kink at just below 0.8 GeV is due to the threshold of the $e^\pm \rho^\mp$ channel.

In addition, to account for the degrading tracking capabilities with increasing distance from the IP, we model the displaced-tracking efficiencies as linear functions of the decay position of the sterile neutrinos in the transverse direction, set to be 0 at $r=80$ cm and 100\% at $r=0$.
This approach has been applied in Refs.~\cite{Dib:2019tuj,Dey:2020juy,Cheung:2021mol,Bertholet:2021hjl,Cheung:2024oxh}.

Finally, we assume 100\% efficiency of tagging the standard-decaying $B$-meson in each signal event.

\section{Simulation and computation}\label{sec:simulation_computation}

We perform Monte-Carlo (MC) simulations in order to determine the detector acceptance to the long-lived sterile neutrinos which includes the fiducial-volume definition as well as the linear displaced-tracking efficiencies, for both the 1DV and 2DV search strategies.

In each simulated event, we first generate the $\Upsilon(4S)$ decay into $B^+ B^-$ at rest, and then boost the $B^+$-meson to the laboratory frame.
We note that with this approach the minor effect of initial state radiation is not included.
We subsequently let the $B^+$-meson decay to $K^+$ and $a$, and let the $a$ decay to a pair of sterile neutrinos.
The intermediate $a$ and the final $N$'s are all boosted to the lab frame.
We thus compute the signal acceptance with input of both the kinematics of the two simulated sterile neutrinos in each event and the detector geometries, in addition to the proper lifetime of the sterile neutrino.

Since detail differs between the 1DV and 2DV search strategies, we elaborate on the computation procedure for the signal-event number $N_S$ in the following two subsections.

\subsection{Computation of $N_S$ in the 1DV signal region}\label{subsec:1DV}

To compute the signal-event number $N_S$ in the 1DV signal region, we use the following expression,
\begin{eqnarray}
    N_S^{\text{1DV}} &=& 2\, N_{B^+B^-}\cdot \text{Br}(B^+\to K^+ a)\cdot \text{Br}(a\to N N)\nonumber\\
        &&\cdot \, \epsilon_{\text{tracks}}^{\text{1DV}} \cdot \epsilon^{\text{PID}}\cdot \text{Br}(N\to \text{signal}), \label{eqn:NS_1DV}
\end{eqnarray}
where $N_{B^+ B^-}$ is the total number of $B^+ B^-$ production events at Belle II, Br$(a\to NN)=100\%$ is the decay branching ratio of the ALP into a pair of the sterile neutrinos, and $\epsilon^{\text{PID}}$ is the efficiency for particle identification (PID) including all final-state particles (except active neutrinos) on the signal-$B$ side.
For $\epsilon^{\text{PID}}$, we assume a flat value of $\epsilon^{\text{PID}}=0.1$.
This is a reasonable and somewhat conservative value chosen based on the realistic estimate for similar displaced final states ($e^+ e^-, \mu^+\mu^-, \pi^+\pi^-, K^+ K^-$) with two DVs presented in Appendix~A of Ref.~\cite{Cheung:2024oxh} and obtained mainly according to fake rates and identification efficiencies at the Belle II detector reported in Ref.~\cite{Belle-II:2018jsg}.
The factor 2 in Eq.~\eqref{eqn:NS_1DV} accounts for the fact that there are two $B$-mesons in each signal event.

The efficiency factor $\epsilon_{\text{tracks}}^{\text{1DV}}$ contains the effect of both the fiducial volume and the displaced-tracking efficiencies, and is estimated with the following formulas:
\begin{eqnarray}
    \epsilon_{\text{tracks}}^{\text{1DV}}&=& \frac{1}{N_{\text{MC}}}\sum_{i=1}^{N_{\text{MC}}}\Big( P_{N_{i,1}}^{\text{1DV}} \cdot (1-P_{N_{i,2}}^{\text{1DV}})\nonumber\\
    &&\quad \quad \quad \quad \quad +P_{N_{i,2}}^{\text{1DV}} \cdot (1-P_{N_{i,1}}^{\text{1DV}})\Big),\\
    P_{N_{i,j}}^{\text{1DV}} &=& \frac{1}{R^{N_{i,j}}}\int_{0}^{80\text{ cm}} L^{\text{1DV}}(r) A(z_{N_{i,j}}) e^{-r/R^{N_{i,j}}}\, dr,\,\,\,\,\,
\end{eqnarray}
where $N_{\text{MC}}$ is the number of the MC-simulated signal events, $R^{N_{i,j}}=\frac{p_T^{N_{i,j}}}{m_N}c\tau_N$ with $j=1,2$ is the boosted decay length of the sterile neutrino $N_{i,j}$ in the transverse direction, and $z_{N_{i,j}}=r \cot{\theta_{N_{i,j}}}$ is the sterile-neutrino decay position in the longitudinal direction determined by the transverse decay position $r$ and the polar angle $\theta_{N_{i,j}}$ of the sterile neutrino in the laboratory frame.
The function $A(z)$ is defined as follows,
\begin{eqnarray}
    A(z) &=& \begin{cases}
        1, -40 < z < 120 \text{ cm}\\
        0, \text{ otherwise,}
    \end{cases}
\end{eqnarray}
and the linear displaced-tracking function is given as
\begin{eqnarray}
    L^{\text{1DV}}(r) &=& \begin{cases}
        1 - \frac{r}{80\text{ cm}}, 10 < r < 80 \text{ cm}\\
        0, \text{ otherwise.}
    \end{cases}
\end{eqnarray}

\subsection{Computation of $N_S$ in the 2DV signal region}\label{subsec:2DV}

In the 2DV signal region, the reconstruction of two DVs is required.
The following formula is used for calculating the signal-event number,
\begin{eqnarray}
    N_S^{\text{2DV}} &=& 2\, N_{B^+B^-}\cdot \text{Br}(B^+\to K^+ a)\cdot \text{Br}(a\to N N)\nonumber\\
        &&\cdot \, \epsilon_{\text{tracks}}^{\text{2DV}}\cdot \epsilon^{\text{PID}}\cdot \Big( \text{Br}(N\to \text{signal} ) \Big)^2,\label{eqn:NS_2DV}
\end{eqnarray}
where the acceptance $\epsilon_{\text{tracks}}^{\text{2DV}}$ is now calculated with
\begin{eqnarray}
    \epsilon_{\text{tracks}}^{\text{2DV}}&=& \frac{1}{N_{\text{MC}}}\sum_{i=1}^{N_{\text{MC}}}P_{N_{i,1}}^{\text{2DV}}\cdot P_{N_{i,2}}^{\text{2DV}},\\
    P_{N_{i,j}}^{\text{2DV}} &=& \frac{1}{R^{N_{i,j}}}\int_{0}^{80\text{ cm}} L^{\text{2DV}}(r) A(z_{N_{i,j}}) e^{-r/R^{N_{i,j}}}\, dr.\,\,\,\,\,\,\,\,\,\,\,\,
\end{eqnarray}
While the $A(z)$ function remains the same as that used in the 1DV signal region, the linear displaced-tracking function is modified to the following expression,
\begin{eqnarray}
    L^{\text{2DV}}(r) &=& \begin{cases}
        1 - \frac{r}{80\text{ cm}}, 1 < r < 80 \text{ cm}\\
        0, \text{ otherwise,}
    \end{cases}
\end{eqnarray}
to take into account the now loosened minimal radial cut.

\begin{figure*}[t]
	\centering
	\includegraphics[width=0.7\textwidth]{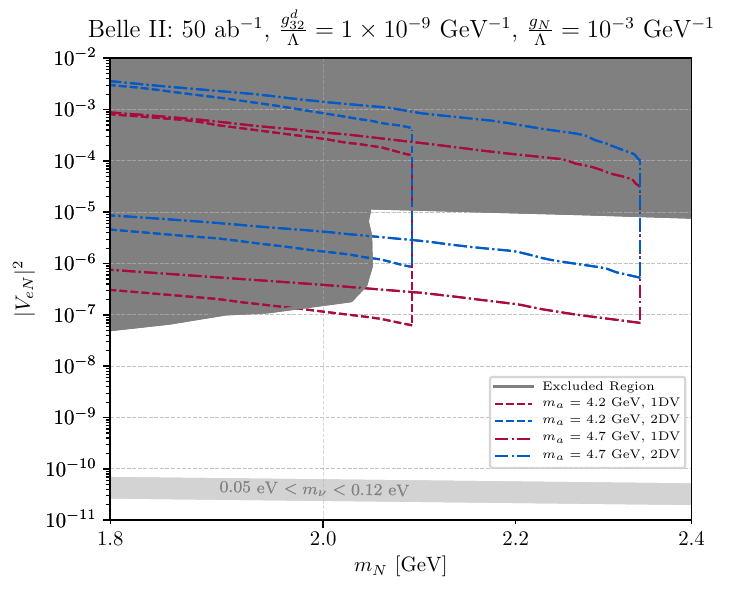}
	\caption{Sensitivity reach of Belle II with the 1DV and 2DV search strategies to the long-lived sterile neutrinos produced in decays of an ALP originating from $B^+\to K^+ a$ decays, shown in the $|V_{eN}|^2$ vs.~$m_N$ plane.
        Here, we have fixed $g_N/\Lambda=10^{-3}$ GeV$^{-1}$, $g^d_{3,2}/\Lambda=1\times 10^{-9}$ GeV$^{-1}$, and $m_a$ at 4.2 GeV (dashed) or 4.7 GeV (dot-dashed).
 The dark-gray area is the currently excluded parameter space for the sterile neutrino in the minimal scenario, extracted from Refs.~\cite{Bryman:2019bjg,CHARM:1985nku,Barouki:2022bkt,CMS:2024ake}, and the light-gray band is the parameter region that can explain the non-zero active-neutrino masses between 0.05 eV and 0.12 eV, under the assumption of the type-I seesaw mechanism.}
 \label{fig:sensitivity_gN1em3}
\end{figure*}

\section{Numerical results}\label{sec:results}

In Fig.~\ref{fig:sensitivity_gN1em3}, we present the sensitivity reach of Belle II with an integrated luminosity of 50 ab$^{-1}$ to the long-lived sterile neutrinos produced in on-shell ALP decays, for $g_N/\Lambda = 10^{-3}$ GeV$^{-1}$.
We show a plot for $m_a$ at 4.2 GeV (dashed) and 4.7 GeV (dot-dashed), respectively, with $g_{3,2}^d/\Lambda$ fixed at $1\times 10^{-9}$ GeV$^{-1}$.
In the plot, we display sensitivity curves of three signal events, corresponding to exclusion bounds at 95\% CL.
The red (blue) curves are for the 1DV (2DV) search strategy.
The 1DV search can always exclude lower values of $|V_{eN}|^2$ than the 2DV search, because of the double exponential suppression present in the 2DV strategy.
Note that we choose to restrict the shown $m_N$ range to be around 2 GeV which is the only region where our proposed searches can probe unexcluded parameter space.

Furthermore, the current bounds on $|V_{eN}|^2$ with respect to $m_N$ are shown in the dark-gray area, obtained at fixed-target, beam-dump, and collider experiments~\cite{Bryman:2019bjg,CHARM:1985nku,Barouki:2022bkt,CMS:2024ake}.
A light-gray band is also overlaid, corresponding to the parameter region that could explain the non-vanishing active-neutrino mass via the type-I seesaw relation $|V_{eN}|^2 \simeq m_{\nu}/m_N$, for $m_{\nu}$ between 0.05 and 0.12 eV.
The lower bound of 0.05 eV is derived from experiments searching for neutrino oscillations~\cite{Canetti:2010aw}, and the upper bound of 0.12 eV stems from cosmological observations~\cite{Planck:2018vyg}.

In Fig.~\ref{fig:sensitivity_gN1em3}, we observe that for heavier sterile neutrinos the Belle II sensitivities to $|V_{eN}|^2$ are stronger.
Consequently, the results for a heavier ALP which allows to probe larger values of $m_N$ shows sensitivity reach to lower values of the mixing angle.
At the strongest point, Belle II can be sensitive to $|V_{eN}|^2$ as low as just below $10^{-7}$ for $m_N$ close to $m_a/2$, with the projected integrated luminosity of 50 ab$^{-1}$ and the 1DV search.
We find that in particular, for $m_N$ above the $D$-meson threshold, the present limits are rather weak compared to those at smaller $m_N$ values, and therefore Belle II with even the 2DV search strategy is already sensitive to values of $|V_{eN}|^2$ about one order of magnitude stronger than the present bounds.

We note that in the large decay-length limit (the lower parts of the bounds shown), the signal-event number is proportional to $\Big(  \frac{g^d_{3,2}}{\Lambda}\Big)^2 |V_{eN}|^2$ (for 1DV) or  $\Big(  \frac{g^d_{3,2}}{\Lambda}\Big)^2 |V_{eN}|^4$ (for 2DV), and therefore, if we, for instance, vary $g^d_{3,2}/\Lambda$ by a factor of 10, the lower reach to $|V_{eN}|^2$ should correspondingly change by a factor of $\frac{1}{100}$ or $\frac{1}{10}$, respectively.

Finally, we comment that although the 2DV search, as expected, has inferior sensitivities compared to the 1DV one, it has the advantage that once signal events are observed, full event reconstruction is possible with certain semi-leptonic final states of the sterile-neutrino decays that have $\mathcal{O}(10\%)$ branching fractions, thus allowing to determine the masses of the sterile neutrino as well as the ALP.

\section{Summary and outlook}\label{sec:conclusions}

Sterile neutrinos can be pair produced in decays of an ALP which originates from rare decays $B^+\to K^+ a$ at $B$-factories such as the ongoing experiment Belle II.
The sterile neutrinos in this mass range are necessarily long-lived, considering the current bounds on their mixing strengths with the active neutrinos.
In this work, we have focused on the signal process, $B^+\to K^+ a, a\to N N$ with an on-shell $a$, and studied the sensitivity reach of Belle II to such long-lived sterile neutrinos.

The long-lived sterile neutrinos can lead to exotic signatures of displaced vertices in the fiducial volume of the detector which we define to be the vertex detector and the tracker.
We have proposed two search strategies, for exactly one DV and two DVs, respectively, stemming from the displaced decays of the sterile neutrinos in each signal event.
In order to determine the detector acceptance within the two signal regions, we have performed MC simulations, taking into account both the defined fiducial volume and displaced-tracking efficiencies.
In particular, to model the deteriorating tracking efficiencies for increasing distances from the IP, we apply functions linear in the transverse distance from the IP.
Moreover, a selected list of signal final states are chosen, in order to ensure that the proposed searches suffer from little to no background contamination that especially originates from photon-conversion processes.

We fix the QFV coupling of the ALP, $g^d_{3,2}/\Lambda$, at $1\times 10^{-9}$ GeV$^{-1}$, considering the latest Belle II measurement results on Br$(B^+\to K^+ \nu \bar{\nu})$ as well as Br$(B^{+/0}\to K^{*+/0} \nu \bar{\nu})$, and assume the other ALP couplings with the quarks are all vanishing.
We take two benchmark ALP masses of 4.2 GeV and 4.7 GeV when showing numerical results.
For the ALP-sterile-neutrino coupling, we consider perturbativity requirement and assume $g_N/\Lambda=10^{-3}$ GeV$^{-1}$, as a benchmark value.
We thus derive the expected sensitivity of Belle II to the active-sterile-neutrino mixing parameter $|V_{eN}|^2$ as functions of the sterile-neutrino mass $m_N$.
For the presentation of the numerical results, we have  focused on the projected integrated luminosity of 50 ab$^{-1}$ with $2.827\times 10^{10}$ $e^- e^+\to B^+ B^-$ events.

We find that with $g_N/\Lambda=10^{-3}$ GeV$^{-1}$, for $m_N$ above the $D$-meson threshold, new parameter space up to two orders of magnitude beyond the current bounds on $|V_{eN}|^2$ can be probed.
If we consider ALP masses below 4 GeV, the proposed searches can be sensitive to sterile-neutrino masses only below the charm-meson threshold, and thus all the sensitive parameter regions have already been excluded.
We note that these results are valid also for other values of $g_N/\Lambda$ as long as the perturbativity requirement is fulfilled and the ALP stays short-lived.

Moreover, the 1DV search strategy can exclude lower values of $|V_{eN}|^2$ than the 2DV one, because the 2DV signal region suffers from the double suppression of the exponential distribution of the sterile-neutrino decay positions.
However, the 2DV search possesses the advantage that if both $N$'s in a signal event undergo semi-leptonic decays it is possible to fully reconstruct the event, allowing to determine the masses of the sterile and the ALP, once the experiment has made a discovery by observing a few such events.

The results of this work could be further strengthened if off-shell ALP contributions are taken into account~\cite{Araki:2024uad} especially for a relatively light ALP, which is beyond the scope of this work.
In addition, one could study the signal process $B^0/B^0_s\to a\to NN$ with the same or similar couplings as those assumed in this work.
In this case, the upper mass reach of $m_N$ can be enhanced to a minor extent since there is no accompanying final-state meson in the signal processes.

Searches similar to those proposed here could be performed at the BaBar~\cite{BaBar:2001yhh} and LHCb~\cite{LHCb:2008vvz,LHCb:2014set} experiments.

\section*{Acknowledgment}

We would like to thank Abner Soffer and Arsenii Titov for useful discussions.
Y.Z.~is supported by the National Natural Science Foundation of China under Grant No.~12475106 and the Fundamental Research Funds for the Central Universities under Grant No.~JZ2023HGTB0222.
W.L.~is supported by National Natural Science Foundation of China (Grant No.~12205153).

\bibliography{bib}

\end{document}